# Pair versus Solo Programming – An Experience Report from a Course on Design of Experiments in Software Engineering

Omar S. Gómez, José L. Batún and Raúl A. Aguilar[1]

[1]Faculty of Mathematics, Autonomous University of Yucatan
Merida, Yucatan 97119, Mexico

**Abstract**
This paper presents an experience report about an experiment that evaluates duration and effort of pair and solo programming. The experiment was performed as part of a course on Design of Experiments (DOE) in Software Engineering (SE) at Autonomous University of Yucatan (UADY). A total of 21 junior student subjects enrolled in the bachelor's degree program in SE participated in the experiment. During the experiment, subjects (7 pairs and 7 solos) wrote two small programs in two sessions. Results show a significant difference (at α=0.1) in favor of pair programming regarding duration (28% decrease), and a significant difference (at α=0.1) in favor of solo programming with respect to effort (30% decrease). With only a difference of 1%, our results regarding duration and effort are practically the same as those reported by Nosek in 1998.

## 1. Introduction

Since the seminal work of Fisher on principles of experimental design [13], the design of experiments (DOE) for obtaining information has been widely used in natural sciences, social sciences and engineering.

When a researcher is designing an experiment, (s)he is interested in analyzing the effect produced in a treatment or intervention that is applied on certain objects or experimental units such as: Persons, plants, animals, etc. SE experiments use to employ persons acting as experimental units, where persons are asked to perform certain tasks that usually constitute a treatment or intervention.

The SE degree program at Autonomous University of Yucatan offers a course on DOE. In this course, students learn to analyze the effect produced in a treatment or intervention by using different types of experimental designs.

As part of this course, during the summer semester 2012 we decided to carry out an experiment; this with the aim of students learn to collect and analyze measures given an experimental design. The experiment selected for the course consisted in analyzing a couple of pair programming aspects.

One of the twelve main practices of extreme programming created by Kent Beck in the late 90s [3, 4] is pair programming. In this practice, two programmers work together on the same task using a computer. One of the programmers (the driver) writes the program whereas the other (the observer) reviews actively the work done by the controller. The observer reviews against possible defects, writes down annotations, or defines strategies for solving any issue that can rise over the task they are working on.

Some experiments have been conducted to study the effect of pair programming [24, 28, 19, 21, 22, 7, 20]. In a general way, these experiments report beneficial effects of applying this practice. Some beneficial effects reported are that it helps to produce shorter programs and helps to implement better designs; programs contain less defects than those written individually, and pairs usually require less time to complete a task than programmers working individually.

Under an academic context, the experiment proposed for the DOE course analyzes the duration and effort needed to write small programs in pairs and individually. The rest of the paper is organized as follows: Section 2 presents the experiment definition. Section 3 describes the design and conduction of the experiment. Section 4 presents the analysis. Section 5 discusses some experiment limitations. In section 6 we discuss the results we found. Finally, in section 7 we present the conclusions and further work.

## 2. Experiment Definition

We use the Goal-Question-Metric approach [2] for defining the experiment. This approach facilitates to identify the object of study, purpose, quality focus, perspective and context of an experiment. We define the experiment as follows:

Study pair and solo programming with the purpose of evaluating possible differences between these two programming types with respect to duration and effort. This study is conducted from the point of view of the researcher under an academic context. This context is composed by juniors students enrolled in a course of DOE where they will write, by pairs or individually, two small programs.

From the experiment definition we derive the following hypotheses:

$H_0 1$: The time required to write a program in pair is equal to the time required to write it individually or: Pair programming = Solo programming regarding time duration.

$H_a 1$: The time required to write a program in pair is different to the time required to write it individually or: Pair programming ≠ Solo programming with respect to time duration.

$H_0 2$: The effort required to write a program in pair is equivalent to the effort required to write it individually or: Pair programming = Solo programming regarding effort.

$H_a 2$: The effort required to write a program in pair is different to the effort required to write it individually or: Pair programming ≠ Solo programing with respect to effort.

## 3. Experiment Design and Conduct

The previous hypotheses will be tested through different measures that we will collect from subjects during the experiment. In a general way, measures belong to two subject groups: Those who perform a task in pairs and those who perform it individually. With these measures, we will perform statistical analyses given an experimental design.

At the beginning of the DOE course, we decided to conduct the experiment at the midterm (semester) in order to students had certain knowledge of DOE and that they had sufficient time to write a report before the semester ended.

The experimental design to use was selected according to the designs listed in the DOE course syllabus. Specifically, we chose the Latin square design because it was scheduled in the course syllabus at midterm, just a few days before the experiment was conducted.

### 3.1 Latin Square Designs

The main features of Latin square designs are that there are two blocking factors. Each treatment is present at each level of the first blocking factor and is also present at each level of the second blocking factor. This design is arranged with an equal number of rows (factor one) and columns (factor two). Treatments are represented by Latin characters symbols where each symbol is present exactly once in each row, and exactly once in each column. An example of the arrangement of this design is shown in Table 1.

Table 1: Latin square design with three treatments

| A | B | C |
|---|---|---|
| B | C | A |
| C | A | B |

In a Latin square design, blocking is used to systematically isolate the undesired source of variation in the comparison among treatments. In this case, pair versus solo programming. As a teaching purpose, we decided to block treatments by program and by tool support. Table 2 shows the arrangement used for the experiment.

Table 2: Latin square design arrangement

| *Program / Tool Support* | *IDE* | *Text Editor* |
|---|---|---|
| Calculator | Solo | Pair |
| Encoder | Pair | Solo |

The program block has two levels: Calculator an encoder whereas tool support block has the levels: IDE (Integrated Development Environment) and text editor. The treatments to examine are: Pair and solo programming.

### 3.2 Subjects, Tasks and Objects

Junior students enrolled in the DOE course participated as subjects in the experiment; in total, for this experiment there were 21 subjects. Most of the subjects were in their third year of the program's degree in SE; the rest of them (three subjects) were in their four year. According of Dreyfus and Dreyfus programming expertise classification [12], we categorized subjects as advanced beginners; subjects have working knowledge of key aspects of Java programming practice.

Subjects were randomized and allocated into two groups: Pair and solo programmers. The experiment was split into two sessions, where in each session subjects wrote a different program. In both sessions we employed the same subjects, so we collected 14 measures with respect to solo programmers (7 solos per session) and 14 measures regarding pair programmers (7 pairs per session). In the first session, subjects that worked individually used NetBeans IDE (as tool support) to write the first program, whereas subjects that worked in pairs used only a text editor. In the second session the tool support was changed, so subjects that before worked individually with the NetBeans IDE, in the second session they worked with a text editor and conversely (See Latin square design arrangement in Table 2).

Before the experiment was conducted, we gave a talk to the students about pair programing. In this talk we explained the main concepts of this programming practice and how it can be used in practice. We also explained how to compile a Java program using only a text editor. Finally, we explained to students how to collect the measures during the experiment sessions. The collection procedure consisted in writing down the time duration that students spent writing a program. They recorded the start and finish time and computed the difference (in minutes).

We selected to small programs that subjects could write, compile, run and test in each session. In the first program (identified as calculator) we asked the subjects to write a calculator that evaluates expressions with decimal numbers, and the operators: Plus (+), minus (-), times (×), divide (/), and prints the result on the screen. In the second program (identified as encoder) we asked the subjects to write a simple encoding-decoding program. Given a specified letter switch the program must be able to encode or decode a line of text.

### 3.3 Conduct

The allotted time for each session was 90 minutes. Both sessions were carried out in one of the computer classroom of the faculty. The first session started almost 30 minutes late because we were waiting for some students to arrive. Once students were complete, we started the session. We gave to subjects some directions and projected on the screen the specification of the program to be written (program calculator). Due to we did not start on time, some subjects did not complete the assignment, so we asked them to pause their work and record the time. Subjects that were working individually we asked them to finish the program at home. At the other hand, subjects that were working in pairs and did not complete the program, we programmed them an extra session on the next day. In this extra session all the remaining pairs completed the program.

The second session started on time; again, we gave to subjects some directions and projected on the screen the second specification (program encoder). In this session all the subjects finished on time. In both sessions programs were verified according to its specification.

### 3.4 Measures

We used the time records of subjects to define the following measures:

**Duration:** It is the elapsed time in minutes to write the program. Before starting the program assignment, subjects wrote down the current time. When they completed the program, they registered the finish time; then we calculate the difference in minutes between start and finish time.

**Effort:** It measures the amount of labor spent to perform a task. It is the total programming effort in person-minutes to write a program. Total effort for a pair is the duration multiplied by two. Tables 3 and 4 show the measures (in minutes) collected for the experiment.

Table 3: Measures collected for duration

| Program / Tool Support | IDE | Text Editor |
|---|---|---|
| Calculator | Solo: 110, 136, 281, 239, 126, 69, 205 | Pair: 256, 184, 114, 59, 37, 89, 135 |
| Encoder | Pair: 70, 48, 88, 85, 43, 39, 56 | Solo: 66, 102, 128, 107, 106, 76, 64 |

Table 4: Measures collected for effort

| Program / Tool Support | IDE | Text Editor |
|---|---|---|
| Calculator | Solo: 110, 136, 281, 239, 126, 69, 205 | Pair: 512, 368, 228, 118, 74, 178, 270 |
| Encoder | Pair: 140, 96, 176, 170, 86, 78, 112 | Solo: 66, 102, 128, 107, 106, 76, 64 |

## 4. Data Analysis

Once we have the measures, we are able to test the hypotheses through statistical inferences. The statistical model associated with a Latin square design is shown in equation (1).

$$y_{ijk} = \mu + \alpha_i + \beta_j + \tau_k + \epsilon_{ijk} \qquad (1)$$

Where $\mu$ is the overall mean, $\alpha_i$ is the block effect common to row $i$, $\beta_j$ is the block effect common to column $j$, $\tau_k$ is the $k$ th treatment effect, and $\epsilon_{ijk}$ is a random error which is assumed to be N(0, $\sigma^2$).

This design uses analysis of variance (ANOVA) to assess the components (overall mean, blocks, treatment and random error) of the model. ANOVA is based on looking at the total variability of the collected measures and the variability partition according to different components. ANOVA provides a statistical test of whether or not the means of several groups are all equal. The null hypothesis is that all groups are simply random samples of the same population. This implies that all treatments have the same effect (perhaps none). Rejecting the null hypothesis implies that different treatments result in altered effects. In this experiment, we have two groups of means (Pair and Solo programming), which are blocked by program and tool support.

### 4.1 Model Assumptions

Before we start to draw any conclusion, we must assess the following model assumptions:
1. All observations are independent (independence)
2. The variance is the same for all observations (homogeneity)
3. The observations within each treatment group have a normal distribution (normality)

The first assumption is addressed by the principle of randomization used in this experimental design; all the measures of one sample are not related to those of the other sample. The second and third assumptions are assessed by using the estimated residuals [6, 16]. To assess homogeneity of variances we use a plot to show a scatter plot of the standardized residuals against the estimated mean values (sometimes called fitted values). We also use the Levene test for homogeneity of variances [17]. The third assumption (normality) is evaluated by using a normal probability plot, and applying the Kolmogorov-Smirnov test for normality [15, 26].

Selecting the duration measure, Fig. 1 shows a scatter plot of the standardized residuals versus fitted values. Violations to the homogeneity variance assumption can be detected with either plot by noting that the variation in the vertical direction seems to differ at different points along the horizontal axis. In this case, Fig. 1 shows a different pattern between the vertical points. Applying the Levene test [17] we get a p-value of 0.0594. Setting an alpha level of 0.05 this test is significant (selecting only two decimal of the p-value with no rounding off), so the assumption of homogeneity is violated.

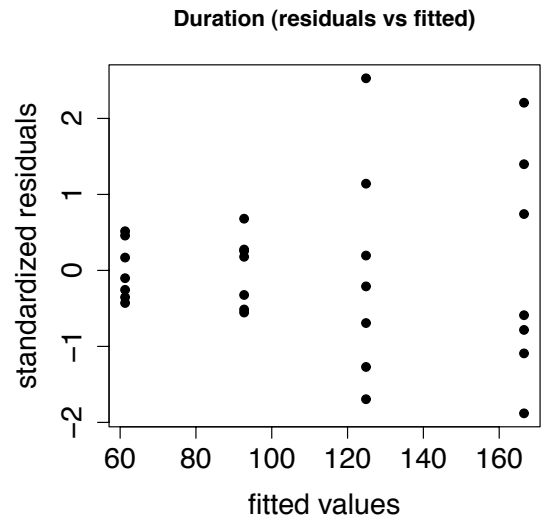

Fig. 1 Scatter plot of standardized residuals vs. fitted values.

Taking a further analysis, we found that the time duration to write the second program was less than the first one. In Fig. 1, the first and second vertical data points correspond to the second program (encoder). Fig. 2 shows the mean time duration to write both programs. To fulfill this assumption, in future experiments we will select programs with similar complexity.

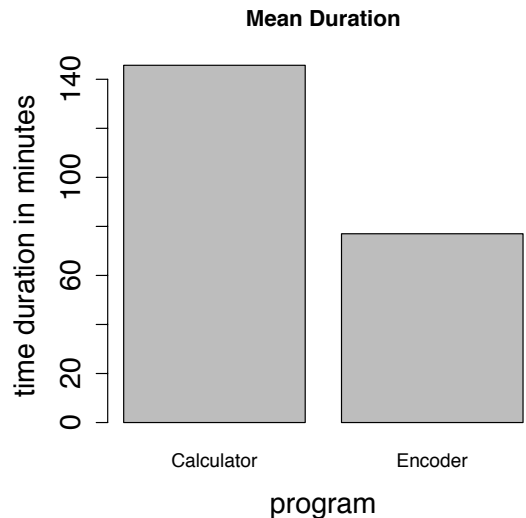

Fig. 2 Mean duration to write a program.

Continuing with the next assumption assessment, Fig. 3 shows a normal probability plot. If points (in this case standardized residuals) fall close to a straight line pattern then residuals are approximately normal. Points that are above the straight line pattern correspond to residuals that are bigger than we might expect for normal data. Points that are below the straight line pattern correspond to residuals that are smaller than we might expect for normal data. Applying the Kolmogorov-Smirnov test for normality [15, 26] we get a p-value of 0.8806; it means that we accept the null hypothesis in favor of normality.

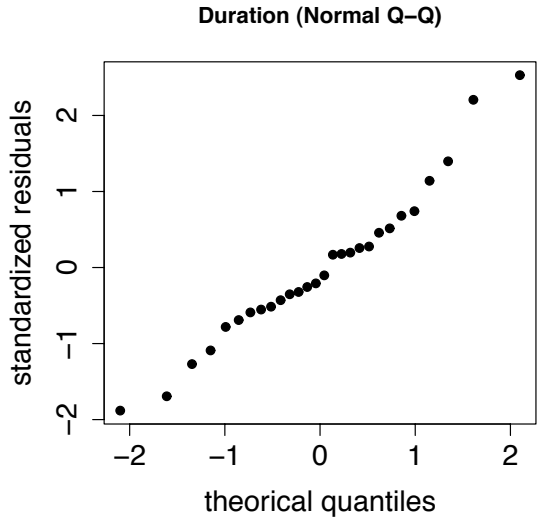

Fig. 3 Normal probability plot.

With respect to the assumptions assessment for effort, we get similar results to those we report regarding duration; performing the Levene test for homogeneity of variances [17] we get a p-value of 0.0241. Setting an alpha level of 0.05 this test is significant. It means that variances are not equal due to differences between programs duration. The Kolmogorov-Smirnov test for normality [15, 26] gives a p-value 0.8059. It means that we accept the null hypothesis in favor of normality.

Due to the experimental design used, another assumption that is worth to assess is the additivity. Experiment designs that implement blocking assume that there is no interaction between the treatment and the block. Under this situation it is told that treatment and block effects are additive [16]. We test this assumption by using the Tukey test for nonadditivity [27]. Table 5 shows the results of this test for the Latin square design used in the experiment.

Table 5: Nonadditivity test results

| *Measure* | *Block* | *F-value* | *p-value* |
|---|---|---|---|
| Duration | Program | 0.0084 | 0.9277 |
| Duration | Tool support | 1.0936 | 0.3061 |
| Effort | Program | 0.0899 | 0.7669 |
| Effort | Tool support | 0.9861 | 0.3306 |

Setting an alpha level of 0.1 (or less), p-values are not significant. It means that experiment results satisfy the assumption of additivity in lack of interaction between treatment and blocks.

## 4.2 Analysis of Variance (ANOVA)

Once model assumptions were assessed, we proceed to perform the ANOVA. Table 6 shows the ANOVA for the duration measure whereas Table 7 shows the ANOVA for effort.

Table 6: ANOVA for duration measure

| *Source* | *Df* | *SS (Type I)* | *MS* | *F-value* | *p-value* |
|---|---|---|---|---|---|
| ProgramBlock | 1 | 33,052 | 33,052 | | |
| ToolSupport Block | 1 | 185 | 185 | | |
| Treatment | 1 | 9,362 | 9,362 | 2.9843 | 0.0969 |
| Residuals | 24 | 75,293 | 3,137 | | |

If we set an alpha level of 0.05 neither treatment (both ANOVA tests) are significant. However setting an alpha level of 0.1 which represents a confidence level of 90% we get significant differences in both treatments. For the first treatment (Table 6) we get a p-value = 0.0969 with respect to duration, whereas we get a p-value = 0.1017 for the second treatment (Table 7). Although this second p-value is slightly greater than 0.1, we also consider it significant.

Table 7: ANOVA for effort measure

| *Source* | *Df* | *SS (Type I)* | *MS* | *F-value* | *p-value* |
|---|---|---|---|---|---|
| ProgramBlock | 1 | 70,702 | 70,702 | | |
| ToolSupport Block | 1 | 4,969 | 4,969 | | |
| Treatment | 1 | 22,346 | 22,346 | 2.8953 | 0.1017 |
| Residuals | 24 | 185,232 | 7,718 | | |

## 4.3 Treatment Comparisons

Taking this alpha level ($\alpha=0.1$) into account, we perform a treatment comparison test (also referred as contrast test) for each measure. Table 8 shows the treatment means, standard error and replications for duration measure whereas Table 9 shows the same information for effort.

Table 8: Treatment means, standard error and replications for duration

| *Treatment* | *Duration (minutes)* | *Std. err* | *Replication* |
|---|---|---|---|
| Solo | 129.6428 | 17.8114 | 14 |
| Pair | 93.0714 | 16.7054 | 14 |

Table 9: Treatment means, standard error and replications for effort

| *Treatment* | *Effort (minutes)* | *Std. err* | *Replication* |
|---|---|---|---|
| Solo | 129.6428 | 17.8114 | 14 |
| Pair | 186.1429 | 33.4108 | 14 |

There are several tests for performing treatment comparisons. These tests help us to analyze pairs of means to assess possible differences between means. Using Scheffé test [21] for treatment comparisons, Table 10 shows the treatment comparison with respect to duration.

Table 10: Comparison with respect to duration

| Comparison | Difference | p-value | LCL (95%) | UCL (95%) |
|---|---|---|---|---|
| Solo-Pair | 36.5714 | 0.0969 | 6.1578 | 66.9850 |

As shown in Table 10, there is a significant difference (at α=0.1) of 36 minutes in favor of pair programming (28% decrease in time). At a confidence interval of 95% this difference ranges between 6 and 66 minutes (4% to 51% decrease in time).

Table 11 shows the treatment comparison with respect to effort. As we see, there is a significant difference (at α=0.1) of 56 minutes in favor of solo programming (30% decrease in effort). At a confidence interval of 95% this difference ranges between 8 and 104 minutes (4% to 55% decrease in effort).

Table 11: Comparison with respect to effort

| Comparison | Difference | p-value | LCL (95%) | UCL (95%) |
|---|---|---|---|---|
| Pair-Solo | 56.5 | 0.1017 | 8.7967 | 104.2032 |

4.4 Effect Size and Power Analysis

Effect size is a measure for quantifying the difference between two data groups. Usually, it is used to indicate the magnitude of a treatment effect. Using the function defined in equation (2) [5], we calculate Cohen's *d* coefficient [10]. This coefficient is used as an effect size estimate for the comparison between two means (in this case Solo and Pair programming). According to Cohen [10], a *d* value between 0.2 and 0.3 represents a small effect size, if it is around 0.5 it is a medium effect size, and an effect size bigger than 0.8 is a large one.

$$d = \sqrt{\frac{F(n_1 + n_2)}{n_1 n_2}} \qquad (2)$$

Using the F-value 2.9843 of the first ANOVA (Table 6) we get an effect size *d* of 0.6529 and an effect size *d* of 0.6431 for the F-value 2.8953 regarding second ANOVA (Table 7). According to Cohen's classification, both effect sizes are considered medium effects. The first effect size is against of solo programming (with respect to duration) whereas the second is against of pair programming (with respect to effort).

Once we have calculated effect sizes, we carry out a power analysis. The power of a statistical test is the probability of rejecting the null hypothesis when it is false. In other words, the power indicates how sensitive is a test to detect an effect in the treatment examined.

Power is equal to 1−β where β is the probability of committing a Type II error [10]. Power analysis can be conducted before or after the experiment is run. When it is performed before, a sample size is estimated with the aim of achieving an adequate power in the statistical test used in the experiment. On the other hand, when the experiment is run, power analysis is used to determine what the power was in the experiment test. We use this second approach to perform power analysis.

Once we know the effect size it is possible to compute the power of a test. In order to determine the power, we use the function *pwr.t.test()* of the R environment [9] which implements power analysis as outlined by Cohen [10]. Given an effect size of 0.6529 (related to duration) and a sample size of n=14 (number of measures in each group; pair and solo programming), and setting a significance level α=0.1; we get a power of 0.51 (51%). Similarly, a power of 0.5 (50%) was obtained with the same sample size and significance level, but replacing the effect size for the value 0.6431 (related to effort).

## 5. Experiment Limitations

Experiments are subject to concerns regarding validity. In this section we discuss experiment limitations based on the four categories of threats to validity described in [11]. Each category has several threats that can negatively impact on the experiment results. We list, both, threats that can impact on this experiment and suggestions for improvements in future versions of this experiment.

5.1 Threats to the Conclusion Validity

These threats concern with issues that affect the ability to draw a correct conclusion about the existence of a relationship between the treatment and the outcome. Next, we describe threats in this category that may have affected our experiment.

Although the experiment results show a moderate power of 50%, results may have been affected by low statistical power. With the aiming of increase the power at 80%, we will perform a power analysis to estimate the sample needed before we conduct replications of this experiment.

Regarding to assumptions of statistical tests, although experiment results satisfy the principle of independence and normality, results may have been affected by lack of variance homogeneity. We have identified the program as a source of variation. With the aiming of reduce variance heterogeneity, in future replications we will use programs with similar complexity.

Another threat that might have affected conclusion validity is with respect to reliability of measures. Although all measures were collected during second session, some measures regarding solo programmers were not collected during first session; it was due to time constraint. In this session subjects that did not finish on time were told to record the time at home. To avoid this threat in future replications we will be careful with managing the time of sessions.

### 5.2 Threats to Internal Validity

These threats concern whether the observed outcomes were due to other factors and not due to the treatment. To avoid these threats, subjects were randomly assigned to the treatments. Latin square design eliminated possible problems with learning effects, boredom or fatigue as the subjects tried different program and tool support. Subjects (pairs and solos) were in the same classroom with equal working conditions, and sitting apart with no interaction.

A possible threat that might have exposed this validity is that subjects knew the experiment, so a competition between pairs and solos could have happened.

### 5.3 Threats to Construct Validity

Construct validity threats concern the relationship between theory and observation. An issue in our experiment that might have affected this validity is that subjects had little or no previous experience with pair programming and they had not programmed with their partners before. These experiment results might be conservative with respect to the effect of pair programming. In subsequent experiment replications, we will reinforce this validity by assigning training programs to pairs.

### 5.4 Threats to External Validity

These threats concern with issues that may limit our ability to generalize the results of the experiment to other contexts, for example generalize it to industry practice. The use of students as subjects instead of practitioners might have exposed this validity. However, as pointed in [8] the use of students as subjects enable us to obtain preliminary evidence to confirm or refute hypotheses that can be tested later in industrial settings.

## 6. Discussion

In this section we discuss some results of other experiments and we contrast them with our results regarding duration and effort.

### 6.1 Duration

The experiment run by Nosek [24] employed 15 practitioners grouped in 5 pairs and 5 solos. Subjects wrote a database script. Results show a decrease of 29% in time duration in favor of pair programming.

Williams et al. [28] used 41 students grouped in 14 pairs and 13 solos. During the experiment, subjects completed four assignments. Authors reported that pairs completed the assignments 40 to 50 percentage faster.

Nawrocki and Wojciechowski [23] employed 16 student subjects (5 pairs and 6 solos). Subjects wrote four programs. Authors did not find differences between pairs and solos.

Lui and Chan [19] used 15 practitioners grouped in 5 pairs and 5 solos. Authors reported 52% decrease in time in favor of pair programming.

Müller [22] used 38 students (14 pairs and 13 solos). Students worked on four programming assignments where tasks were decomposed into implementation, quality assurance and the whole task. Author reported that pairs spent 7% more time working on the whole task, however this difference is not significant.

Arisholm et al. [1] used 295 practitioners grouped in 98 pairs and 99 solos. Subjects performed several change tasks on two alternative systems with different degrees of complexity. Authors reported 8% decrease in favor of pairs.

In contrast, the results reported in this paper infer a significant (at $\alpha=0.1$) 28% decrease in time (in favor of pairs) and an effect size $d=0.65$. With respect to duration, our results reinforce those reported in [24].

### 6.2 Effort

This measure is not present in all of the experiments previously discussed, so we compute it (doubling the time duration of pairs) only in the cases where data is available.

According to Nosek data [24] we observe a decrease in effort of 29% in favor of solo programming. Conversely, data of Lui and Chan [19] indicate a decrease of 4% in favor of pairs. Finally, Arisholm et al. [1] Report an increase in effort of 84% (against of pairs).

In contrast, the results reported in this paper infer a significant (at α=0.1) 30% decrease in effort (in favor of solos), and an effect size d=0.64. Our results, again, reinforce those calculated in [24].

## 7. Conclusions and Further Work

This paper presented a controlled experiment that was run as part of a university course in DOE. The aim of the experiment was to evaluate pair versus solo programming with respect to duration and effort. Subjects who jointly wrote the program assignments took less time (28%) than subjects who worked individually. Conversely subjects grouped in pairs spent more effort (30%) than those who worked individually. These results are very close to those reported in [24].

With the aiming of striving towards better research practices in SE [18] we reported all the collected measures. This data will help other researchers to verify or re-analyze [14] the experiment results presented in this work. This data can also be used to accumulate and consolidate a body of knowledge about pair programing.

We are planning to conduct future replications of this experiment to get more insight about the effect of pair programming. Although we did not observe interactions between treatment and blocks, we plan to use another experimental design to assess possible interactions.

## References

[1] E. Arisholm, H. Gallis, T. Dybå, and D. I. Sjøberg. Evaluating pair programming with respect to system complexity and programmer expertise. IEEE Transactions on Software Engineering, 33(2):65–86, 2007.
[2] V. Basili, G. Caldiera, and H. Rombach. Goal question metric paradigm. Encyclopedia of Software Eng, pages 528–532, 1994. John Wiley & Sons.
[3] K. Beck. Embracing change with extreme programming. Computer, 32(10):70–77, 1999.
[4] K. Beck. Extreme programming explained: embrace change . Addison-Wesley Longman Publishing Co., Inc., Boston, MA, USA, 2000.
[5] M. Borenstein. The handbook of research synthesis and meta analysis. Chapter: Effect sizes for continuous data, pages 279–293. Russell Sage Foundation, New York, USA, 2009.
[6] G. E. P. Box, W. G. Hunter, J. S. Hunter, and W. G. Hunter. Statistics for Experimenters: An Introduction to Design, Data Analysis, and Model Building. John Wiley & Sons, June 1978.
[7] G. Canfora, A. Cimitile, F. Garcia, M. Piattini, and C. A. Visaggio. Evaluating performances of pair designing in industry. Journal of Systems and Software, 80(8):1317 – 1327, 2007.
[8] J. Carver, L. Jaccheri, S. Morasca, and F. Shull. Issues in using students in empirical studies in software engineering education. In METRICS '03: Proceedings of the 9th International Symposium on Software Metrics, page 239, Washington, DC, USA, 2003. IEEE Computer Society.
[9] S. Champely. pwr: Basic functions for power analysis , 2012. R package version 1.1.1.
[10] J. Cohen. Statistical power analysis for the behavioral sciences . L. Erlbaum Associates, Hillsdale, NJ, 1988.
[11] T. Cook and D. Campbell. The design and conduct of quasi-experiments and true experiments in field settings. Rand McNally, Chicago, 1976.
[12] H. L. Dreyfus and S. Dreyfus. Mind over Machine. The Power of Human Intuition and Expertise in the Era of the Computer . Basil Blackwell, New York, 1986.
[13] R. A. Fisher. The Design of Experiments. Oliver & Boyd, Edimburgh, 1935.
[14] O. S. Gómez, N. Juristo, and S. Vegas. Replication, reproduction and re-analysis: Three ways for verifying experimental findings. In International Workshop on Replication in Empirical Software Engineering Research (RESER'2010) , Cape Town, South Africa, May 2010.
[15] A. N. Kolmogorov. Sulla determinazione empirica di una legge di distribuzione. Giornale dell'Istituto Italiano degli Attuari, 4:83–91, 1933.
[16] R. Kuehl. Design of Experiments: Statistical Principles of Research Design and Analysis. Duxbury Thomson Learning, California, USA. second ed. edition, 2000.
[17] H. Levene. Robust tests for equality of variances. In I. Olkin, editor, Contributions to probability and statistics . Stanford Univ. Press. Palo Alto, CA, 1960.
[18] P. Louridas and G. Gousios. A note on rigour and replicability. SIGSOFT Softw. Eng. Notes, 37(5):1–4, Sept. 2012.
[19] K. M. Lui and K. C. C. Chan. When does a pair outperform two individuals? In Proceedings of the 4th international conference on Extreme programming and agile processes in software engineering, XP'03, pages 225–233, Berlin, Heidelberg, 2003. Springer-Verlag.
[20] K. M. Lui, K. C. C. Chan, and J. Nosek. The effect of pairs in program design tasks. IEEE Trans. Softw. Eng., 34(2):197–211, Mar. 2008.
[21] C. McDowell, L. Werner, H. E. Bullock, and J. Fernald. The impact of pair programming on student performance, perception and persistence. In Proceedings of the 25th International Conference on Software Engineering , ICSE '03, pages 602–607, Washington, DC, USA, 2003. IEEE Computer Society.
[22] M. M. Müller. Two controlled experiments concerning the comparison of pair programming to peer review. Journal of Systems and Software, 78(2):166 – 179, 2005.
[23] J. Nawrocki and A. Wojciechowski. Experimental evaluation of pair programming. In Proceedings of the 12th European Software Control and Metrics Conference, pages 269–276, London, April 2001.
[24] J. T. Nosek. The case for collaborative programming. Commun. ACM , 41(3):105–108, Mar. 1998.
[25] H. Scheffé. A method for judging all contrasts in the analysis of variance. Biometrika , 40(1/2):87–104, 1953.

[26] N. V. Smirnov. Table for estimating the goodness of fit of empirical distributions. Ann. Math. Stat., 19:279–281, 1948.
[27] J. W. Tukey. One degree of freedom for non-additivity. Biometrics, 5(3):pp. 232–242, 1949.
[28] L. Williams, R. Kessler, W. Cunningham, and R. Jeffries. Strengthening the case for pair programming. Software, IEEE, 17(4):19 –25, jul/aug 2000.

**Omar S. Gómez** received a BS degree in Computing from the University of Guadalajara (UdG), and a MS degree in Software Engineering from the Center for Mathematical Research (CIMAT), both in Mexico. Recently, he received a PhD degree in Software and Systems from the Technical University of Madrid (UPM). Currently he is a full time professor of Software Engineering at Mathematics Faculty of the Autonomous University of Yucatan (UADY). His main research interests include: Experimentation in software engineering, software process improvement and software architectures.

**José L. Batún** received a BS degree in Mathematics from the Autonomous University of Yucatan (UADY). He received a MS degree and a PhD degree in Probability and Statistics, both, from the Center for Mathematical Research (CIMAT) in Guanajuato, Mexico. He is currently full time professor of Statistics at Mathematics Faculty of the Autonomous University of Yucatan (UADY). His research interests include: Multivariate statistical models, copulas, survival analysis, time series and their applications.

**Raúl A. Aguilar** was born in Telchac Pueblo, Mexico, in 1971. He received the BS degree in Computer Science from the Autonomous University of Yucatan (UADY) and a PhD degree (PhD European mention) at the Technical University of Madrid (UPM), Spain. Currently he is full time professor of software engineering at Mathematics Faculty of the Autonomous University of Yucatan (UADY). His main research interests include: Software engineering and computer science applied to education.